\documentclass[%
 reprint,
 amsmath,amssymb,
 aps,
]{revtex4-1}

\usepackage{graphicx}
\usepackage{dcolumn}
\usepackage{bm}

\begin{document}

\preprint{APS/123-QED}

\title{Inflationary scale, reheating Scale and pre-BBN Cosmology with scalar fields}

\author{Alessandro Di Marco}
\email{alessandro.di.marco@roma2.infn.it}
\author{Gianfranco Pradisi}
\author{Paolo Cabella}

\affiliation{%
University of Rome - Tor Vergata, Via della Ricerca Scientifica 1\\
INFN Sezione di Roma “Tor Vergata”, via della Ricerca Scientifica 1, 00133 Roma, Italy}%


\begin{abstract}

In this paper, we discuss the constraints on the reheating temperature supposing an early post-reheating cosmological phase
dominated by one or more simple scalar fields produced from inflaton decay and decoupled from matter and radiation.
In addition, we explore the combined effects of the reheating and non-standard scalar field phases on the 
inflationary number of $e$-foldings.

\end{abstract}

\pacs{Valid PACS appear here}
\maketitle


\section{Introduction}

Before the Hot Big Bang (HBB) epoch, our Universe likely  experienced an early quantum gravity phase (at the so called Planck scale) in which gravitational, strong, weak and electromagnetic interactions were unified in a single fundamental force \cite{1}. 
Due to expansion and cooling, at lower (GUT) scales the gravitational interaction decoupled and  
the Universe entered an hypothetical phase where matter and radiation can be described in terms of
a Grand-Unified gauge theory \cite{2}.  According to the inflationary paradigm, at a scale $M_{inf}$ ($<10^{16}$ GeV), after the spontaneous symmetry breaking to $SU (3)\times SU (2)\times U (1)$ (the gauge group of the Standard Model of particle physics), cosmological inflation is supposed to have taken place, in order to make the Universe almost flat, isotropic and homogeneous on large astronomical scales \cite{3}.
In the simplest version, the inflationary mechanism was driven by a  scalar field, called inflaton, minimally coupled to gravity and probing an almost flat region (a false vacuum) of the corresponding effective scalar potential.  At the end of inflation, where the potential steepens, the inflaton field falls in the global minimum of the potential, oscillates, decays and ``reheats" the Universe
(see \cite{4} for detailed studies on the mechanism and \cite{5} for general constraints), giving rise to the standard HBB evolution characterized by an initial radiation-dominated phase.  However, this last step is not necessarily the unique possible scenario.  
Indeed, there is of course room for a peculiar evolution in the history of the Universe immediately after the reheating. 
In particular, the expansion of the Universe could have been submitted to additional phases where, for instance, it was driven by one (or more) new simple scalar species, before the radiation-dominated era and, especially, well before the Big Bang Nucleosinthesys (BBN).
Additional scalar fields, not necessarily directly interacting with the Standard Model degrees of freedom, are quite common in superstring theory with branes. They typically parametrize the brane positions along  directions internal to the extra-dimensions transverse to the branes.  Since their energy density exhibits a modified dilution law, they can give rise to a non-standard post-reheating phase.  
Scenarios of this type have been recently introduced to study modification on relics abundances and decay rates of dark matter \cite{6,7}, as well as to study enhancements in the inflationary number of $e$-foldings \cite{8,9}. In this paper, we consider non-standard cosmologies inspired by string theory orientifold models
\cite{10} with, generically, multiple sterile scalar fields entering a non-standard post-reheating phase and we analyze in details the constraints put on the reheating temperature by the additional fields.  
As a consequence, we can derive more stringent model independent predictions about the number of $e$-folds during inflation.
The paper is organized as follows. In Sec. II, we derive general expressions for the energy density in the case of non-standard post-reheating cosmological evolution, given by one or more scalar fields. 
In Sec. III, we discuss how the features of the new species affect the reheating scale.  In particular, we derive an upper limit to the reheating temperature. In Sec. IV, we study the relation between reheating and postinflationary scalar fields and we calculate the inflationary number of $e$-foldings, also constrained by the maximum reheating temperature.  In Sec. V, we add our conclusions and some discussions.  In the Appendixes, we show numerical examples of the consequences of the variation in the number of $e$-foldings on the inflationary predictions of $n_s$ and $r$, for various selected inflaton potentials.
In this manuscript we use the particle natural units $c=\hbar=1$, unless otherwise stated.

\section{postinflationary scalar fields and cosmology} 

The cosmological history of early Universe immediately after the reheating should be characterized by a 
radiation-dominated era. In that phase, the corresponding evolution is well described by 
\begin{eqnarray}
H^2(T)\simeq\frac{1}{3M^2_{Pl}}\rho_{rad}(T), \quad \rho_{rad}(T)=\frac{\pi^2}{30}g_{E}(T) T^4 ,
\label{matdom}\end{eqnarray}
where 
$H$ denotes the Hubble rate, $M_{Pl}$ is the reduced Planck Mass, $\rho_{rad}$ is the radiation energy density and $T$ indicates the temperature
scale of the universe at a given (radiation-dominated) epoch. Finally, $g_{E}$ is the effective number of relativistic degrees of freedom turning out to be
\begin{equation}
g_{E}(T)=\sum_{b} g_{b}\left( \frac{T_b}{T}\right)^4 + \frac{7}{8}\sum_{f} g_{f}\left(\frac{T_f}{T}\right)^4, 
\end{equation}
where $b$ and $f$ label contributions from bosonic and fermionic degrees of freedom, respectively, and $T_b$ and $T_f$ indicate the corresponding temperatures.
In this Section, we would like to analyze a  modification of the evolution of the early Universe after the reheating phase, realized through the presence of a set of scalar fields $\phi_i (i=1,...,k)$.  They are assumed to dominate at different time scales until radiation becomes the most relevant component, well before the BBN era \cite{6}.
The last assumption is crucial in order to not spoil the theoretical successes related to the prediction of light element abundances
(see \cite{7}).  
Therefore, the total energy density after  the inflaton decay can be assumed to be
\begin{equation}\label{eqn:totalenergydensity}
\rho(T)=\rho_{rad}(T) + \sum_{i=1}^k \rho_{\phi_i}(T).
\end{equation}
We introduce the scalar fields in such a way that, for $i>j$, $\rho_{\phi_{i}}$ hierarchically dominates at higher temperatures over $\rho_{\phi_{j}}$ when the temperature decreases.
All the scalar fields, supposed to be completely decoupled from each other and from matter and radiation fields, can be described as perfect fluids diluting faster than radiation.
In this respect, the dynamics is encoded in 
\begin{equation}
\dot{\rho}_{\phi_i} + 3H\rho_{\phi_i}( 1 + w_i ) = 0 , 
\end{equation}
where $w_i=w_{\phi_i}$ is the equation of state (EoS) parameter of the field $i$.
Integrating this equation one finds 
\begin{equation}
\rho_{\phi_i}(T)=\rho_{\phi_i}(T_i)\left(\frac{a(T_i)}{a(T)} \right)^{4+n_i}, \quad n_i=3w_i -1
\end{equation}
where the index $n_i$, the ``dilution" coefficient, is understood to satisfy the conditions
\begin{equation}
n_i>0, \quad n_i<n_{i+1}.
\end{equation}
$T_i$ can be conveniently identified with the transition temperature  
at which the contribution of the energy density of $\phi_i$ becomes subdominant with respect to the one of $\phi_{i-1}$.
In other words, the scalar fields are such that  
\begin{eqnarray}
\rho_{\phi_i}>\rho_{\phi_{i-1}} \mbox{ for } T>T_i\\
\rho_{\phi_i}=\rho_{\phi_{i-1}} \mbox{ for } T=T_i\\
\rho_{\phi_i}<\rho_{\phi_{i-1}} \mbox{ for } T<T_i.
\end{eqnarray} 
Using the conservation of the ``comoving" entropy density
\begin{equation}
g_S(T)a^3(T)T^3=g_S(T_i)a^3(T_i)T^3_i ,
\end{equation}
being $g_S$, defined by
\begin{equation}
g_{S}(T)=\sum_{b} g_{b}\left( \frac{T_b}{T}\right)^3 + \frac{7}{8}\sum_{f} g_{f}\left(\frac{T_f}{T}\right)^3,
\end{equation}
the effective number of relativistic degrees of freedom associated with entropy, the energy density of the various fields at a temperature $T$ can be expressed in terms of the transition temperatures $T_i$ \cite{6,9}
\begin{equation}
\rho_{\phi_i}(T)=\rho_{\phi_i}(T_i)\left( \frac{g_S(T)}{g_S(T_i)}\right)^{\frac{4+n_i}{3}}\left(\frac{T}{T_i}\right)^{4+n_i}.
\label{ratioofrhophi}\end{equation}
For the first scalar field $\phi_1$, by definition, the transition temperature is such that its energy density is identical to the one of the radiation fluid, so that
\begin{equation}
\rho_{\phi_1}(T_1)=\rho_{rad}(T_1)=\frac{\pi^2}{30}g_{E}(T_1) T_1^4.
\end{equation}
The second scalar field $\phi_2$ is subdominant compared to $\phi_1$ below the temperature $T_2$.  Using Eq.\eqref{ratioofrhophi}
and observing that $T_2$ is the transition temperature at which $\rho_{\phi_2}(T_2)=\rho_{\phi_{1}}(T_2)$, one gets
\begin{equation}\label{eqn: energy phi2}
\rho_{\phi_2}(T)= \rho_{\phi_1}(T_1)\left(  \frac{T_2 g_S^{1/3}(T_2)}{T_1 g_S^{1/3}(T_1)} \right)^{4+n_1}  \left(  \frac{T g_S^{1/3}(T)}{T_2 g_S^{1/3}(T_2)}  \right)^{4+n_2}
\end{equation}
This equation tells us that the energy density of the scalar field $\phi_2$ depends on the ratio between the two scales $T_1$ and $T_2$,
where the $\phi_1$-dominance occurs. 
In the same way, we can derive the analogous expression for the other scalar fields $\phi_i$. The general expression for the energy density carried by $\phi_i$ turns out to be
\begin{widetext}
\begin{equation}\label{eqn: general result}
\rho_{\phi_i}(T)=\rho_{\phi_1}(T_1)\prod_{j=1}^{i-1}\left(\frac{T_{j+1} g_S^{1/3}(T_{j+1})}{T_j g_S^{1/3}(T_j)}\right)^{4+n_j}
\left(\frac{T g_S^{1/3}(T)}{T_i g_S^{1/3}(T_i)}\right)^{4+n_i}, \quad i\ge 2.
\end{equation}
\end{widetext}
Inserted in Eq. \eqref{eqn:totalenergydensity}, the previous expressions provide the total energy density dominating the expansion of the Universe after the standard reheating phase, up to the beginning of the radiation-dominated epoch. In particular, the Hubble rate acquires the compact form 
\begin{equation}\label{eqn: hubble rate evo}
H^2(T)\simeq\frac{1}{3 M^2_{Pl}} \rho_{\phi_1}(T_1)\sum_{i=1}^{k} f_i(n_i,T,T_1,...,T_i),
\end{equation}
where $f_i$ can be extracted by the previous equations.

\section{Nature of the scalar fields and reheating temperature}

In the previous section, we have discussed a modified post-reheating scenario, where several component species in the form of non-interacting (decoupled) scalar fields, are added to the relativistic plasma.  Even if we do not specify their nature, it should be underlined that these kind of components are quite common both in scalar modifications of General Relativity and in theories with extra dimensions. In particular, in orientifold superstring models compactified to four dimensions and equipped with D-branes, the presence of additional scalars is an almost ubiquitous phenomenon \cite{10}. Indeed, the D-brane action is the sum of a DBI term  and a Wess-Zumino term, generalizations of the familiar mass and charge terms of a particle action. The dynamical fluctuations of the D-branes in the transverse directions correspond to degrees of freedom that are described by scalar fields. Their coupling to the four-dimensional metric is induced on D-branes by the embedding inside the ten-dimensional space-time, and gives rise typically to a warp factor depending on the internal coordinates and to additional couplings entering the DBI action (disformal terms, see \cite{9} and references therein).  The most important point, however, is that these scalar fields always interact with the inflaton, that can thus decay into them and the remaining components of the standard reheating fluid after inflation.  In this paper, we neglect the interactions of the scalar fields with matter longitudinal to the D-branes.
From Eq.\eqref{eqn: general result} we can easily argue that $\rho(T)$ increases with temperature, reaching the maximum value at $T=T_{reh}$.  For instance, in the case of a single additional scalar field $\phi_1$, with a transition-to-radiation temperature $T_1$, one has
\begin{equation}
\rho_{\phi_1}(T)=\rho_{\phi_1}(T_1)\left( \frac{g_S(T)}{g_S(T_1)}\right)^{\frac{4+n_1}{3}}\left(\frac{T}{T_1}\right)^{4+n_1}.
\end{equation}
It is clear that there must be an upper bound to this energy density for $T=T_{reh}$. 
At this stage, whatever the nature of $\phi_1$ is, the energy density cannot assume arbitrary values, since it is at least  limited by the presence of the Planck scale, $M_{Pl}$.
In other words, we have to introduce a maximum scale $M$ (with $M\le M_{Pl}$) such that 
\begin{equation}
\rho_{\phi_1}(T_{reh})\leq M^4 , 
\end{equation}
corresponding to an upper limit to the production scale of $\phi_1$.  As a consequence, it turns out also to be an upper limit to the reheating temperature, once we set the scale $M$.
Since $T_1$ is the transition-to-radiation temperature, i.e.
\begin{equation}
\rho_{\phi_1}(T_1)=\rho_{rad}(T_1) ,
\end{equation}
using 
\begin{equation}\label{eqn: g_relation}
g_{E}(T)\sim g_{S}(T)\sim 100 \mbox{ for } T>T_{QCD} 
\end{equation}
(where $T_{QCD}>150$ MeV is the QCD phase transition scale), 
the reheating temperature must satisfy the condition
\begin{equation}\label{eqn: temp bound}
T_{reh}\leq \alpha_1 M \left(\frac{T_1}{M} \right)^{\frac{n_1}{4+n_1}}, \quad \alpha_1=\left(\frac{30}{\pi^2 g_E}\right)^{\frac{1}{4+n_1}}
\end{equation}
with a resulting upper limit 
\begin{equation}\label{eqn: templimit}
T^{max}_{reh}=\alpha_1 M \left(\frac{T_1}{M} \right)^{\frac{n_1}{4+n_1}}.
\end{equation}

\begin{figure}[htbp]
\centering
\includegraphics[width=8.5cm, height=6cm]{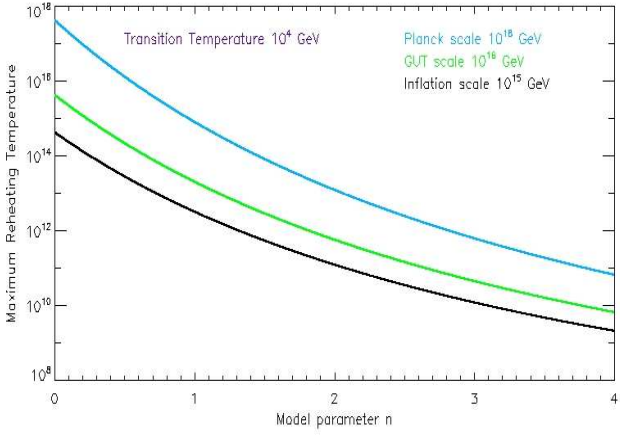}
\caption{The maximum reheating temperature $T^{max}_{reh}$ as a function of the parameter $n$ for $T_1=10^4$ GeV.
The maximum reheating temperature becomes larger as $M$ increases,  while it decreases with $n$.}
\label{fig: 1}
\end{figure}

\begin{figure}[htbp]
\centering
\includegraphics[width=8.5cm, height=6cm]{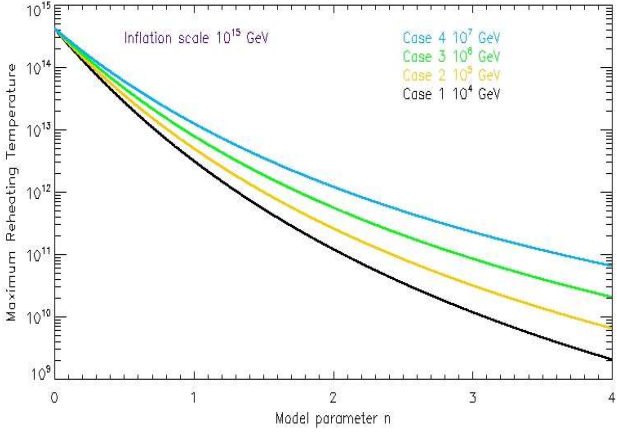}
\caption{The maximum reheating temperature $T_{reh}^{max}$ as a function of the parameter $n$ for different values of the transition-to-radiation
temperature and an inflationary scale $M_{inf}\sim 10^{15}$ GeV.
The maximum reheating temperature becomes larger as the scale $T_1$ increases.}
\label{fig: 2}
\end{figure}

In general, the scale $M$ could be the Planck scale $M_{Pl}$ but also, for instance, the GUT scale $M_{GUT}$ or a lower scale of the order of the string scale, $M_{s}$, that is unconstrained in orientifolds \cite{10,11}. 
In particular, if the field $\phi_1$ is supposed to be produced by the inflaton decay or during the reheating phase, we can also assume $M=M_{inf}$,
where $M_{inf}$ is the inflationary scale.
It is interesting to analyze how the upper limit on $T_{reh}$ varies  with the scale $M$.  In Fig.($\ref{fig: 1}$) we plot the behaviour of  $T^{max}_{reh}$ as a function of the model parameter $n=n_1$ for given values of the scale $M$. The transition-to-radiation temperature is chosen to be $T_1\sim 10^4$ GeV.
As expected, the maximum reheating temperature is larger for larger values of $M$, 
while it decreases with the model 
parameter $n$.
The region below each curve representing $T^{max}_{reh}(n)$ describes the possible reheating temperatures compatible with the chosen bound $M$.
For example, for $n=2$ and $M=M_{Pl}$, we might have $T_{reh}\le 10^{13}$ GeV, while for $M=M_{inf}$ we might only have $T_{reh}\le 10^{11}$ GeV.
For $n=4$, a reheating temperature of the order of $10^9$ GeV is compatible with $M=M_{inf}$ and {\it a fortiori} with Planckian or GUT bounds.  In Fig.($\ref{fig: 2}$) we fix the scale $M$ to the inflationary scale ($\sim 10^{15}$ GeV) and plot the behaviour of $T^{max}_{reh}(n)$ for different values of the transition-to-radiation temperature.  It happens that $T^{max}_{reh}(n)$ becomes smaller and smaller, for a fixed $n$, as the transition temperature decreases. For example, with $n=2$ and $T_1\sim 10^7$ GeV, we get $T_{reh}\le 10^{12}$ GeV, while with $n=4$ $T_{reh}\le 10^{11}$. 
We postpone the discussion of the case with more scalar fields to Section V.

\section{Inflationary $e$-foldings, Reheating and pre-BBN scalar fields}

\begin{figure}[htbp]
\centering
\includegraphics[width=8.5cm, height=6cm]{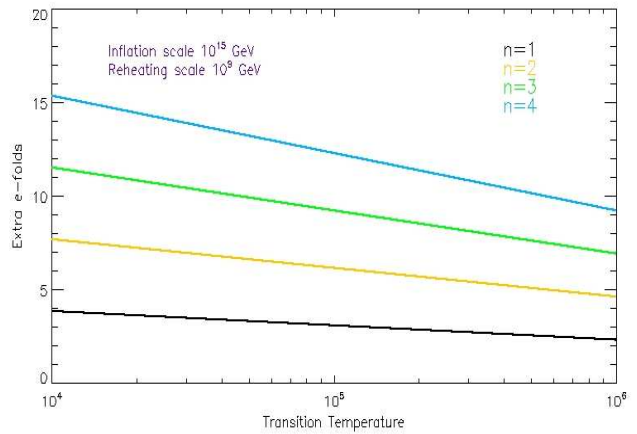}
\caption{Number of extra $e$-folds with $w_{reh}=0$ and $T_{reh}\sim 10^9$ GeV. We have chosen this temperature because it is compatible with all values of $n$ from 1 to 4 and with all the transition temperatures $T_1>10^4$ GeV, as seen in Sec.II.}
\label{fig: 3}
\end{figure}

\begin{figure}[htbp]
\centering
\includegraphics[width=8.5cm, height=6cm]{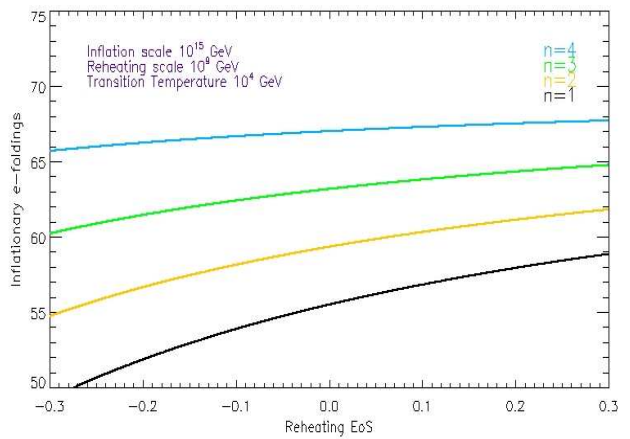}
\caption{The inflationary number of $e$-folds $N_*$ in a nonstandard postreheating cosmology as a function of $w_{reh}$. $N_*$ increases with the value of the EoS parameter.  The growth of $N_*$ is also decreasing with $n$.}
\label{fig: 4}
\end{figure}

In the case of 
standard post-reheating radiation-dominated Universe, the inflationary number of $e$-foldings $N_*$ has been calculated and used in many works \cite{5}.  In this Section, we would like to discuss how this number changes in the presence of 
a non-standard postinflationary scenario. As shown in \cite{9},
in the non-standard case $N_*$ acquires an additional $e$-folds term $\Delta N(\phi_i,T_{reh})$, that depends on the reheating temperature and on the features of the additional decoupled scalar fields discussed in Section II.  Thus, we may write
\begin{align}\label{eqn: efolds}
N_*&= \xi_* -\frac{1}{3(1+w_{reh})}\ln\left(\frac{\rho_{end}}{\rho_{reh}}\right)\\ \nonumber
&+ \frac{1}{4}\ln\left(\frac{V^2_*}{M^4_p \rho_{reh}}\right) + \Delta N(\phi_i,T_{reh}) , 
\end{align}
where $\rho_{end}$ is the energy density at the end of inflation, $\rho_{reh}$ is the energy density when the reheating is completely realized, $w_{reh}$
is the mean value of the EoS parameter of the reheating fluid,
while $V_*=M_{inf}^4$ is the inflationary energy density. In Eq. \eqref{eqn: efolds} 
\begin{equation}
\xi_* = -\ln\left(\frac{k_*}{a_0 H_0}\right) + \ln\left(\frac{T_0}{H_0}\right) + c \ ,
\end{equation}
with
\begin{equation}
c=- \frac{1}{12} \ln g_{reh} + \frac{1}{4}\ln\left(\frac{1}{9}\right) + \ln\left(\frac{43}{11}\right)^{\frac{1}{3}}\left(\frac{\pi^2}{30}\right)^{\frac{1}{4}} ,
\end{equation}
where $k_*$ is the pivot scale for testing the cosmological parameters, $a_0$ and $H_0$ are the scale factor and the Hubble rate
at the current epoch, respectively, $T_0$ is the CMB photon temperature while $g_{reh}$ denotes the effective number of relativistic degrees of freedom at the end of reheating (we are using  $g_E(T_{reh})=g_S(T_{reh})=g_{reh}$ because of Eq.($\ref{eqn: g_relation}$)).
Assuming $k_*=0.002$ Mpc$^{-1}$, $H_0=1.75\times 10^{-42}$ GeV, $T_0=2.3\times 10^{-13}$ GeV and $g_{reh}\sim 100$, we get 
$\xi_*\sim 64$ and $c\sim 0.77$.
The additional term comes out to be
\begin{equation}
\Delta N(\phi_i,T_{reh})=\frac{1}{4}\ln\eta(n_i,T_i,T_{reh}), 
\end{equation}
where $\eta$ is the ratio of the total energy density to the energy density of radiation at the reheating temperature,
\begin{equation}
\eta= 1 + \frac{\sum_i \rho_{\phi_i}(T_{reh})}{\rho_{rad}(T_{reh})}.
\end{equation}
Using Eq.($\ref{eqn: general result}$) and expressing the radiation energy density in terms of $T_1$
\begin{equation}
\rho_{rad}(T_{reh})= \rho_{rad}(T_1) \frac{g_E(T_{reh})}{g_E(T_1)}\left(\frac{T_{reh}}{T_1}\right)^4
\end{equation}
we can write
\begin{widetext}
\begin{equation}\label{eqn: eta_general}
\eta = 1 + \frac{g_E(T_1)}{g_E(T_{reh})}\left(\frac{T_1}{T_{reh}}\right)^4 
\biggl\{\left[\frac{T_{reh} g_S^{1/3}(T_{reh})}{T_{1} g_S^{1/3}(T_{1})}\right]^{4+n_1} + \sum_{i=2}^k \prod_{j=1}^{i-1}  \left[\frac{T_{reh} g_S^{1/3}(T_{reh})}{T_{i} g_S^{1/3}(T_{i})}\right]^{4+n_i} \ \left[\frac{T_{j+1} g_S^{1/3}(T_{j+1})}{T_{j} g_S^{1/3}(T_{j})}\right]^{4+n_j}\biggl\}
.
\end{equation}
\end{widetext}
It should be noticed that the more scalar fields we have, the larger the parameter $\eta$ is. 
Moreover, $N_*$ is inflationary-model dependent due to the presence of the potential function in the second and third contributions of Eq.($\ref{eqn: efolds}$). 
However, by assuming $\rho_{end}\sim M_{inf}^4$, converting $\rho_{reh}$ in $T_{reh}$ and neglecting some small numerical factors, $N_*$ can also be written as 
\begin{align}\label{eqn: efolds ind}
N_*&\sim \xi_* - \frac{1-3w_{reh}}{3(1+w_{reh})}\ln\left(\frac{M_{inf}}{T_{reh}}\right) \\ \nonumber
&+ \ln\left(\frac{M_{inf}}{M_{Pl}}\right)  + \frac{1}{3(1+w_{reh})}\ln\eta .
\end{align}
We can distinguish three main contributions. The first 
\begin{equation}
A(w_{reh},T_{reh})=\frac{1-3w_{reh}}{3(1+w_{reh})}\ln\frac{M_{inf}}{T_{reh}}
\label{eq:reheatingcontr}\end{equation}
is entirely related to the reheating phase, the second involves the ratio between the Planck scale and the inflationary scale while the last one is due to the fraction of energy carried by the scalar fields, namely to the $\eta$ factor.
Let us provide a simple example considering a single scalar field post-reheating dominance.
By using Eq.($\ref{eqn: g_relation}$), the general expression Eq.($\ref{eqn: eta_general}$) turns out to be
\begin{equation}\label{eqn: eta1}
\eta = 1 + \left(\frac{T_1}{T_{reh}}\right)^4\left(\frac{T_{reh}}{T_1}\right)^{4+n_1}\simeq \left(\frac{T_{reh}}{T_1}\right)^{n_1} ,
\end{equation}
and therefore
\begin{equation}\label{eqn: eta_term}
\Delta N(\phi_1,T_{reh})=\frac{n_1}{3(1+w_{reh})}\ln\left(\frac{T_{reh}}{T_1}\right)
\end{equation}
that, for the trivial $w_{reh}=0$ case, results into 
\begin{equation}
\Delta N(\phi_1,T_{reh})=\frac{n_1}{3}\ln\left(\frac{T_{reh}}{T_1}\right).
\end{equation}
The reheating and the $\eta$ terms are strongly correlated.  
Indeed, in Section II we have shown that the reheating temperature is constrained by an upper bound dependent on a scale $M$, by the transition-to-radiation temperature $T_1$ and by the dilution coefficient $n=n_1$.  As a consequence, we have a lower bound on the reheating contribution in Eq.\eqref{eq:reheatingcontr}.
Using the bound in Eq.($\ref{eqn: temp bound}$), we get
\begin{equation}\label{eqn: reh_constraint}
A(w_{reh},T_{reh})\ge\frac{1-3w_{reh}}{3(1+w_{reh})}\ln\frac{M_{inf}}{\alpha_{1}M}\left(\frac{M}{T_1}\right)^{\frac{n_1}{4+n_1}}.
\end{equation}
In Fig.($\ref{fig: 3}$) we report the quantity $\Delta N(\phi_1,T_{reh})$ as a function of the transition-to-radiation temperature for some 
values of $n_1$, assuming an equation of state $w_{reh}=0$ and\\
a reheating temperature $T_{reh}\sim 10^9$ GeV.
Let us take a\\
look
that for $n=4$ and $T_1\sim 10^4$ GeV, we can easily\\
extract 
more than $15$ extra $e$-folds, while for a larger $T_1\sim 10^6$ GeV we would have $\Delta N\sim 9$.
In Fig.($\ref{fig: 4}$) we plot the complete result for the variable $N_*$ as a function of the reheating equation of state parameter $w_{reh}$ for $n=1,2,3,4$, assuming $T_1\sim 10^4$ GeV.
In general, the value of $N_*$ increases with $w_{reh}$, as expected  by the expression in Eq.($\ref{eqn: efolds ind}$).
For $n=2$ and $w_{reh}=0$, we get $N_*\sim 59$, while $N_*\sim 67$ for $n=4$ and $w_{reh}=0$.  
In the next section we briefly discuss the multifield cases.
The obtained results have non-trivial consequences on the theoretical predictions of the underlying inflationary models. 
The reason is that one usually infers the values of the two main inflationary parameters, the scalar spectral index $n_s$ and the tensor-to-scalar ratio $r$, assuming an $N_*$ in the range between 50 and 60.
Therefore, if we considered a different $N_*$ we could have new predictions to compare with the current experimental bounds \cite{12}.
In the Appendixes we will briefly examine how the non-trivial values of $N_*$ affect some paradigmatic inflationary models.

\section{Summary and Discussion}

Inflation should have taken place at very high-energy scales.
The accelerated expansion was followed by a ``reheat" stage that produced the Standard Model radiation fluid and the observable large comoving entropy of the Universe.  However, available data do not guarantee that the mentioned scenario is the correct one.
For instance, a viable alternative is to have one (or more) additional field(s) that dominates the energy budget of the Universe at different phases after the reheating epoch.  In particular, many authors have recently considered the inclusion of new scalar fields (quintessence, scalar field decoupled from matter and radiation
or even scalar fields or moduli coupled to gravity) to approach some problems related to dark matter relics abundances or to the number of  inflationary $e$-folds (see \cite{6,7,8,9} and references therein).
The relation between scalar decaying particle and black hole formation in GUT cosmology was also studied in the past \cite{13}.
In general, nonstandard cosmological histories before the BBN are interesting possibilities whose signatures could be tested in the near future,
for istance by gravity-waves experiments (see \cite{14} for details).

In this paper, extending the approach of \cite{9}, we have described a post-reheating era dominated by a collection of simple scalar fields $\phi_i, (i=1,...,k)$ completely decoupled from Standard Model matter and radiation. Their presence is described in terms of perfect fluids with energy densities that scale as $\rho_i\sim a^{-(4+n_i)}$, $n_i>0$. Each $\phi_i$ dominates at different times. In particular, $\phi_1$ is the field connecting the non-standard part of the post-reheating phase to the radiation-dominated era.  
In this scenario, it is mandatory to assume that the transition to radiation occurs well before the BBN, in order not to ruin the theoretical predictions about light element abundances \cite{6}.
In Eqs. \eqref{eqn:totalenergydensity} and \eqref{eqn: general result} the general expressions related to the total energy density and to the energy density of a single field $\phi_i$ have been derived, with the proviso of absence of entropy variation.  
The changes in the Hubble rate during the multifield driven evolution are regulated by Eqs. \eqref{eqn: general result} and \eqref{eqn: hubble rate evo}.  
In Sec. III, we 
observed that the energy density after reheating must be at most Planckian.  
As a consequence, there exists an upper limit to the reheating temperature, as shown in Eq. ($\ref{eqn: templimit}$) and illustrated in Fig.($\ref{fig: 1}$) for different choices of the limiting scale and a transition temperature to HBB $\sim 10^4$ GeV.  The upper bound depends on $T_1$ and also on the indices $n_i$, as shown in Fig.($\ref{fig: 2}$).  

Let us take a closer look to the multifield case, already mentioned in Sec. III.  Of course, the upper bound on $T_{reh}$ is always present, but it depends on the intermediate temperatures $T_i$.  For instance, in the presence of two scalar fields, with $\phi_2$ dominating at higher temperature $T>T_2$ on $\phi_1$, the condition becomes $\rho_{\phi_2}\leq M^4$ at $T=T_{reh}$. As a consequence, one gets
\begin{equation}
T_{reh}<\alpha_2 M \left( \frac{T_1^{n_1} T_2^{n_2-n_1}}{M^{n_2}} \right)^{\frac{1}{4+n_2}}
\label{eq:boundwithtwo}\end{equation}
where $\alpha_2=\left(30/\pi^2 g_E\right)^{1/4+n_2}$  (by assumption $n_2>n_1$). Using, for instance, $n_1=1$,  $n_2=2$, $T_1\sim 10^4$ GeV and $T_2\sim 10^6$ GeV one has $T_{reh}<10^{12}$ GeV for $M<M_{inf}$, while $T_{reh}<10^{13.7}$ GeV for $M<M_{Pl}$.  Note that in the first case the value of $T_{reh}^{max}$ is very close to the one found in the presence of a single scalar field with $n_1=2$ at a transition-to-radiation temperature $T_1\sim 10^7$ GeV (see Sec.II).  The upper bound can obviously be computed for any number $k$ of scalar fields and it turns out to depend on $2k$ parameters, the $T_i$ temperatures and the $n_i$ dilution coefficients. 

A nonstandard cosmological epoch after reheating gives also rise to an extra term in the general expression of the inflationary number of $e$-foldings, $N_*$ (see Eqs. \eqref{eqn: efolds} and \eqref{eqn: efolds ind}).  The upper bound on the energy density of the $k$-th scalar field leads to an additional constraint on the contribution to $N_*$ coming from the reheating phase, as shown in  Eq.($\ref{eqn: reh_constraint}$).
As a result, we found the possibility of having an inflationary number of $e$-foldings well beyond $60$, as shown in Fig.($\ref{fig: 4}$).  The higher is the number of scalar fields, the larger is the correction $\Delta N$ to $N_*$, since the ratio of the total energy density to the radiation density at $T_{reh}$ is larger. 
For instance, with two scalar fields and using Eq.($\ref{eqn: g_relation}$), Eq.($\ref{eqn: eta_general}$) provides
\begin{widetext}
\begin{equation}
\eta\simeq 1 + \left(\frac{T_1}{T_{reh}}\right)^4\left(\frac{T_{reh}}{T_1}\right)^{4+n_1} + \left(\frac{T_1}{T_{reh}}\right)^4\left(\frac{T_2}{T_1}\right)^{4+n_1}
\left(\frac{T_{reh}}{T_2}\right)^{4+n_2}.
\end{equation}
\end{widetext} 
By choosing the same data as after Eq. \eqref{eq:boundwithtwo} and $w_{reh}=0$, $T_{reh}\sim10^{13}$ GeV, one gets $\eta\sim 10^{16}$, $\Delta N(\phi_1,\phi_2)\sim 12$ and $N_*\sim 70$.
As expected, a nonstandard post-reheating phase produces a variety of enhancements in the inflationary number of $e$-foldings, depending on the the number of additional scalar fields and on the details of their dilution properties.  
Enhancements affect the theoretical predictions of the inflationary models, mainly in the bottom right portion of the familiar $(n_s,r)$ plane.
In \cite{9}, Maharana and Zavala have studied the functions $n_s(N_*)$ and $r(N_*)$. 
In Appendixes A and B, we report some results for typical classes of inflationary models, extending the range of parameters provided in \cite{9}. 
We deserve an extended analysis to a future publication \cite{15}.

\begin{acknowledgments}
This work was supported in part by the MIUR-PRIN Contract 2015MP2CX4 ``Non-perturbative Aspects Of Gauge Theories And String". P.C. would also like to thank the company ``L'isola che non c'\`e S.r.l" for the support.
\end{acknowledgments}

\appendix

\section{Monomial Potentials}

The first class of inflationary models we are going to analyze are those characterized by single monomial potentials of the form
\begin{equation}
V(\varphi)=\lambda_p\varphi^p, \quad \lambda_p=M^4_{inf} M_{Pl}^{-p}.
\end{equation}
In this class of models the inflaton field, in order to drive inflation, must exhibit a super-Planckian variation $\Delta\phi>M_{Pl}$.  Historically, the most known scenarios are the ones with $p=2$ and $p=4$, that were introduced by Linde \cite{3}.
Monomial potentials naturally occur also in superstring compactifications, where they are called ``axion monodromy'' \cite{16}.  In these models, one uses a cover of the compactification manifold, with branes wrapping suitable internal cycles.  As a result, even though the manifold is compact, the wrapping of branes around certain cycles weakly breaks the original shift symmetry,  allowing for closed-string axions with super-Planckian excursions and the suppression of dangerous higher-dimensional  operators. 
The involved inflationary potentials come out precisely of the form $V(\varphi)\sim \varphi^p$ with $p=2/5,2/3,1$ or $4/3$.
The slow-roll parameters give rise to standard theoretical predictions for the spectral index and the tensor-to-scalar ratio in terms of the inflationary number of $e$-foldings:
\begin{equation}
n_s\sim 1-\frac{p+2}{2 N_*}, \quad r=\frac{4p}{N_*}.
\end{equation}
It should be noticed that both $n_s$ and $r$ depend on the model parameter $p$.
In Tab. I and in Tab. II we report the theoretical predictions for some scenarios related to monomial potentials, assuming two possible non-standard post-reheating data.
\begin{table}[htbp]
\begin{ruledtabular}
\begin{tabular}{lcdr}
\textrm{Model parameter $p$}&  \textrm{$n_s(N_*)$}&\multicolumn{1}{c}\textrm{$r(N_*)$}\\
\colrule                      
Axion model $p=2/5$                      & 0.9821                          & 0.0238  \\                       
Axion model $p=2/3$                      & 0.9801                          & 0.0398 \\  
Axion model $p=1$                        & 0.9776                          & 0.0597 \\
Axion model $p=4/3$                      & 0.9751                          & 0.0796   \\
Linde model $p=2$                        & 0.9701                          & 0.1194 \\
Linde model $p=4$                        & 0.9552                          & 0.2388 \\
\end{tabular}
\end{ruledtabular}
\caption{Inflationary predictions for monomial potentials in nonstandard postreheating cosmology. We assume a single scalar field with $T_1\sim 10^4$ GeV, $n_1=4$ and $T_{reh}\sim 10^9$ GeV, giving  $N_*=67$.
}
\label{tab: table1}
\end{table}

\begin{table}[htbp]
\begin{ruledtabular}
\begin{tabular}{lcdr}
\textrm{Model parameter $p$}&  \textrm{$n_s(N_*)$}&\multicolumn{1}{c}\textrm{$r(N_*)$}\\
\colrule                       
Axion model $p=2/5$                       & 0.9830                           & 0.0229  \\                       
Axion model $p=2/3$                       & 0.9810                           & 0.0381 \\  
Axion model $p=1$                         & 0.9785                           & 0.0571 \\
Axion model $p=4/3$                       & 0.9762                           & 0.0762   \\
Linde model $p=2$                         & 0.9714                           & 0.1143 \\
Linde model $p=4$                         & 0.9571                           & 0.2286 \\
\end{tabular}
\end{ruledtabular}
\caption{Inflationary predictions for monomial potentials in nonstandard postreheating cosmology. We assume two scalar fields with $T_1\sim 10^4$ GeV, $n_1=1$, $T_2\sim 10^6$ GeV, $n_2=2$ and $T_{reh}\sim10^{13}$ GeV, giving $N_*=70$.}
\label{tab: table2}
\end{table}

\section{Exponential Potentials}

The second class of models we would like to consider is that of  exponential potentials of the form
\begin{equation}
V(\varphi)\sim M^4_{inf}\left(1-e^{-b\varphi}\right) , \quad b=\sqrt{\frac{2}{3\alpha}} ,
\label{exppot}\end{equation}
where $\alpha$ is a free parameter. These potentials arise in many contexts. Important examples are the well known Starobinsky model ($\alpha=1$), the Goncharov-Linde model ($\alpha=1/9$) and the Higgs Inflation model ($\alpha=\sqrt{2/3}$)\cite{17}.  More recently, the so called $\alpha$-attractor models of inflation \cite{18} have also been considered, that fall in the same class of Eq. \eqref{exppot}. 
Furthermore, other very interesting examples come out in superstring-inspired scenarios, like K\"ahler Moduli Inflation, Poly-instanton Inflation and Fiber Inflation \cite{19}.
At first order, the theoretical predictions of this class of models result
\begin{equation}
n_s\sim 1-\frac{2}{N_*}, \quad r\sim \frac{12\alpha}{N^2_*}.
\end{equation}
In this case, the scalar spectral index does not depend on the value of $\alpha$. Therefore, for $N_*=67$ one has $n_s\sim 0.9701$ while for $N_*=70$, $n_s=0.9714$, independently on  $\alpha$. On the contrary, the tensor-to-scalar ratio depends on $\alpha$ as shown in Tab. III, where we report its values for some choices of the parameters.
\begin{table}[htbp]
\begin{ruledtabular}
\begin{tabular}{lcdr}
\textrm{Model parameter $\alpha$}&       \textrm{$r(N_1)$}&\multicolumn{1}{c}\textrm{$r(N_2)$}\\
\colrule
Starobinsky $\alpha=1$                       & $2.7\times 10^{-3}$            & 2.4\times 10^{-3} \\                       
Fiber Inflation $\alpha=2$                   & $5.3\times 10^{-3}$            & 4.9\times 10^{-3}  \\                       
Goncharov-Linde $\alpha=1/9$                 & $2.9\times 10^{-4}$            & 2.7\times 10^{-4} \\  
Poly-instanton $\alpha=3\times 10^{-3}$      & $8.0\times 10^{-6}$            & 7.3\times 10^{-6}  \\
K\"ahler Moduli $\alpha=3\times 10^{-8}$     & $8.0\times 10^{-11}$           & 7.3\times 10^{-11} \\
\end{tabular}
\end{ruledtabular}
\caption{The tensor-to-scalar ratio for some exponential potential models depending on $\alpha$.
In the first column, we assume $N_1=67$ (related to the case with a single scalar field and $T_1\sim 10^4$ GeV, $n_1=4$, $T_{reh}\sim 10^9$ GeV). In the second column we assume $N_2=70$ (related to a pair of scalar fields characterized by $T_1\sim 10^4$ GeV, $n_1=1$, $T_2\sim 10^6$ GeV, $n_2=2$ and $T_{reh}\sim10^{13}$ GeV.)}
\label{tab: table3}
\end{table}

\nocite{*}

\bibliography{apssamp}

\end{document}